\documentclass[aps,pre,twocolumn,groupedaddress,showpacs]{revtex4-1}
\usepackage{amsmath,amssymb,subfigure,fancybox,txfonts,bm,color,wrapfig}
\usepackage{siunitx,numprint}
\usepackage[dvipdfmx]{graphicx}
\begin{document}
\title{ 
Power-law fluctuations near critical point in semiconductor lasers with delayed feedback
}

\def\tr#1{\mathord{\mathopen{{\vphantom{#1}}^t}#1}} 
\newcommand{\vc}{\mathbf}
\newcommand{\gvc}[1]{\mbox{\boldmath $#1$}}
\newcommand{\emf}[1]{{\gtfamily \bfseries #1}}
\newcommand{\fracd}[2]{\frac{\displaystyle #1}{\displaystyle #2}}
\newcommand{\ave}[1]{\left< #1 \right>}
\newcommand{\red}[1]{\textcolor{red}{#1}}
\newcommand{\blue}[1]{\textcolor{blue}{#1}}
\newcommand{\green}[1]{\textcolor[rgb]{0,0.6,0}{#1}}

\newcommand{\del}[2]{\frac{\partial #1}{\partial #2}}
\newcommand{\dev}[2]{\frac{\text{d} #1}{\text{d}#2}}
\newcommand{\mdev}[3]{\frac{\text{d}^{#3} #1}{\text{d}#2^{#3}}}
\newcommand{\pdev}[2]{{\text{d} #1}/{\text{d}#2}}
\newcommand{\intd}[1]{\text{d} {#1}}
\newcommand{\eng}[1]{#1}
 \newcommand{\jpn}[1]{}
\newcommand{\subti}[1]{\begin{itemize} \item {\bf #1} \end{itemize}}
\newcommand{\niyama}[1]{\textcolor[named]{Red}{(新山) #1}}
\newcommand{\suna}[1]{\textcolor[named]{Red}{(SUNADA) #1}}

\author{Tomoaki Niiyama$^{1}$}
\email{niyama@se.kanazawa-u.ac.jp}
\author{Satoshi Sunada$^{1,2}$}

\affiliation{
${}^1$Institute of Science and Engineering, Kanazawa University, 
Kakuma-machi, Kanazawa, Ishikawa 920-1192, Japan
\\
${}^2$Japan Science and Technology Agency (JST), PRESTO, 4-1-8 Honcho,
 Kawaguchi, Saitama 332-0012, Japan
}
\date{\today}

\begin{abstract}
Since the analogy between laser oscillation and second-order phase transition
was indicated in the 1970s,
dynamical fluctuations on lasing threshold inherent in critical phenomena 
have gained significant interest. 
Here, we numerically and experimentally 
demonstrate that a semiconductor laser subject to 
delayed optical feedback can exhibit unusual large intensity fluctuations
characterized by power-law distributions.
Such an intensity fluctuation consists of distinct intermittent bursts 
of light intensity, whose peak values attain tens of times  
the intensity of the maximum gain mode.
This burst behavior emerges when a laser with a long time delay 
(over $\SI{100}{ns}$)
and an optimal feedback strength operates around the lasing threshold.
The intensity and waiting time statistics follow power-law-like distributions.
We also report on the experimental results that
suggested power-law intensity dynamics in a semiconductor laser
with delayed feedback.
The reason for the emergence of power-law behavior in the laser system
is discussed from the perspective of 
nonequilibrium-critical behavior in a slowly driven-dissipative system
known as self-organized criticality.
\end{abstract}

\maketitle

\section{Introduction}
\label{sec:introduction}

Laser oscillation has gained interest in the areas of theoretical physics
as well as engineering application
since the analogy between laser oscillation and second-order phase transition 
was indicated by DeGiorgio and Scully and Graham and Haken
\cite{DeGiorgio1970LaserPT,Graham1970PT2ndLaser}.
The analogy to the second-order phase transition
has inspired significant interest not just 
in the transition of a macroscopic quantity,
i.e., the onset of laser oscillation,
but also its fluctuation 
around the lasing threshold, which is regarded as a critical point
\cite{Rice1994PhotonStatisticsQEDLaser,
Kuo2012CriticalityPTinVCSELaser,Basak2016LargeFlucutationLasingThreshold,
Wang2017ThresholdDynamicsMesoLaser,Wang2020PhotonBurstLaser}.
However, despite many studies on this fluctuation of light intensity,
scale-invariant fluctuations expected from the context of 
equilibrium critical phenomena have not been reported.

The scale-invariant feature mentioned here implies that
variable $s$ such as magnitude, duration, or waiting time
of an energy-releasing event follows a {\it power-law distribution}:
\begin{align}
P(s) \propto s^{-\beta},
\label{eq:power-law}
\end{align}
where $\beta$ denotes the characteristic exponent.
Power-law distributions, which are generalized as L\'{e}vy's stable
distributions, have a scale-invariant feature because
their mean, variance, and higher order of moments cannot be determined
mathematically under a certain condition of the exponent 
and the upper bound of $s$ \cite{Sornette2006SOC}.
Thus, 
dissipation phenomena characterized by the statistical distribution provide
substantially large fluctuation and intermittency in their energy dissipation.
Such a statistical feature can be found in various natural phenomena 
and physical systems,
e.g., solar flares \cite{aschwanden2011SOC}, 
earthquakes \cite{Omori1894OmoriLaw,Utsu1969UtsuLaw}, 
plasticity \cite{Miguel2001Intermittent-di,Niiyama2015ICPMD}, 
sandpiles \cite{Bak1987SOC,Bak1988SOC}, 
magnetization \cite{Durin2000BarkhausenAvalanche}, 
and superconductors \cite{Field1995SuperConductVortexAvalanche,SOC1998Jensen}.

This scale-invariant feature can be observed in equilibrium-critical phenomena:
for instance, the area of the magnetized domain in the Ising model.
Laser systems differ from those that exhibit equilibrium-critical phenomena 
in many aspects such that laser is a kind of nonequilibrium systems.
One of the most significant differences is the degrees of freedom of the system.
Laser dynamics consists of only a few dynamical variables,
such as the electric field and carrier density \cite{Uchida2012ChaoticLaser},
whereas equilibrium-critical systems have many degrees of freedom.
However, it is well known that delayed feedbacks provide 
high dimensionality to the laser dynamics
 \cite{Ikeda1987HighDimChaosDelayedLaser,Arecchi1992DOFofDelaySystem,Giacomelli1996DOFofDelayedSystem,Uchida2012ChaoticLaser,Erneux2017TimeDelayDynamics}.
Generally, delay systems can be represented as spatially expanded systems,
where numerous independent elements,
the so-called virtual nodes that correspond to many degrees of freedom,
interact with themselves in a time domain.
Hence, delayed feedback can provide multiplicity of the processes 
and their interactions to the system through high dimensionality.

In addition to the high dimensionality, 
the delayed feedback also provides a wide variety of dynamical behaviors 
with laser systems
\cite{Uchida2012ChaoticLaser,Ohtsubo2012semiconductor,Erneux2017TimeDelayDynamics}.
For instance, external light injection with a time delay can cause
regular sequential pulsation in a short timescale, the so-called
rapid pulse packages (RPPs) \cite{Heil2001LK-RPP},
and irregular behaviors, known as optical turbulence,
under a strong optical feedback condition \cite{Ikeda1980OptTurbulence}.
Low pump-current conditions near the lasing threshold
yield a complex fluctuation, called low-frequency fluctuations (LFF),
consisting of sub-nanosecond intensity pulses
and regular intensity drops with a longer timescale
\cite{Fischer1996LFF,Heil1998LFFmap,Ohtsubo2012semiconductor}.
Temporal fluctuations of intermittent pulsation
in similar low pumping-power conditions
have been investigated and analyzed in terms of intermittent dynamics
and extreme events \cite{Reinoso2013CavityLaserEE,Boscos2013DelayedOpticsEE,Choi2016CavityLaserEE,Bosco2017RandNumGen,Barbosa2019StatisticsLaserChaos}.
In addition, the existence of virtual nodes and their interplay 
have been used for delay-based computing 
\cite{Appeltant2011MLtoDelayedSystem,Larger2017ReservoirDelayedSystem}. 

This high dimensionality and complexity originating from delayed feedback
may provide a laser system scale-invariant fluctuations
near the lasing threshold.
In this study, we show the statistical distributions and their features 
in the intensity dynamics of
semiconductor lasers with delayed feedback
by conducting numerical simulations and supplementary experiments.
In Section \ref{sec:simulation}, we demonstrate through numerical simulations
of the Lang--Kobayashi model 
that a semiconductor laser with long and strong delayed feedback
exhibits intermittent intensity bursts characterized by power-law distributions.
Then, the condition to obtain a remarkable power-law feature is 
elucidated based on the simulation results.
In Sec.~\ref{sec:experiment},
we report the experimental results, which partly support the numerical results.
Finally, in Sec.~\ref{sec:discussion}, 
we discuss the emergence of the power-law behavior 
for the proposed laser system
from the viewpoint of the conditions to achieve 
criticality in slowly driven-dissipative systems.

\section{Numerical simulation}
\label{sec:simulation}

\subsection{Numerical simulation setup}

In the present study, 
we employ the Lang--Kobayashi model represented 
by the complex electric field $\hat{E}(t)$ and 
carrier density $N(t)$ at time $t$
for the simulation of a semiconductor laser with a single delayed feedback
\cite{Lang1980LangKobayashiEq,Uchida2012ChaoticLaser}:
\begin{align}
 \dev{\hat{E}}{t} &= 
\frac{1+ i \alpha}{2} 
\left[ G_N \left\{ N(t) - N_0  \right\} - \frac{1}{\tau_p} \right]
\hat{E}(t)
+ \kappa \hat{E}(t - \tau) e^{- i \omega_0 \tau},
\label{eq:LK-complex-E}
\end{align}
\begin{align}
\dev{N}{t} &= J - \frac{N(t)}{\tau_s} - G_0 \left\{ N(t) - N_0 \right\} 
| \hat{E}(t)|^2,
\label{eq:LK-complex-N}
\end{align}
\begin{align}
G_N &= \frac{ G_0  }{ 1+\epsilon |\hat{E}(t)|^2 },
\label{eq:LK-G_N}
\end{align}
where $\alpha$, $G_0$, $N_0$, $\tau_p$, $\tau_s$, $\omega_0$, and $\epsilon$
are the linewidth enhancement factor, gain coefficient, 
carrier density at transparency, photon lifetime, carrier lifetime,
optical angular frequency, and gain saturation parameter, respectively.
Here, $J$, $\kappa$, and $\tau$ represent the pump current,
feedback strength, and external round-trip time corresponding to the delay time,
respectively.
We used the following set of parameters for the simulations:
$\alpha = 5$, $ t_p   = \SI{1.927e-12}{\second}$,
$ t_s   = \SI{2.04e-09}{\second}$, $ N_0 = \SI{1.4E+24}{1/m^3}$,
$\omega_0 = \SI{ 1.226e+15 }{s^{-1}}$ and $G_0 = \SI{1E-12}{m^3/s}$.
The gain saturation parameter is set as zero except for
the simulations in Appendix~\ref{sec:saturation}. 
The fourth-order Runge--Kutta method 
was used to numerically integrate the Lang--Kobayashi model.
The timestep for the integration was set to $1$ femtosecond.

A normalized pump current defined as $j = J / J_{th}$ 
is used in the simulations, where 
$J_{th}$ is the threshold value of 
the solitary oscillation mode, $J_{th} = N_{th}/\tau_s$, and
$N_{th} = N_0 + 1/(\tau_p G_0)$.
Notably, the laser can oscillate below the threshold current $j=1$
because some external cavity modes that originate from the delayed feedback
have lower thresholds than the threshold of the solitary mode.
We also normalize the laser intensity, $I = |\hat{E}|^2$, 
by that of the maximum gain mode,
\begin{align}
 I_{ex} &= \frac{1}{G_N} \frac{\tau_p}{\tau_s} 
\frac{ (j-1) N_{th} G_N + 2 \kappa }{1 - 2 \tau_p \kappa},
\label{eq:I_max}
\end{align}
where the maximum gain mode is the most efficient oscillation mode,
with the lowest lasing threshold,
among the external-cavity modes \cite{Uchida2012ChaoticLaser}.
The derivation of the equation is described in Appendix \ref{sec:ext_mode}.

As initial conditions of the electric field and carrier density,
those of the maximum gain mode were applied:
$\text{Re}[\hat{E}(0)] = \text{Im}[\hat{E}(0)] 
= \sqrt{I_{ex}}$, $N(0) = N_{ex}$, 
where 
$
N_{ex} = N_{th} - \left(2 \kappa/G_N \right)
$.
For the initial condition of the delayed electric field
[$\hat{E}(t)$ for $-\tau \le t < 0$],
a random electric field, $\hat{E}(0)$ with $1\%$ random noise generated by 
uniform random numbers, was used.
We confirmed that the initial condition does not affect 
the dynamical behaviors as long as a sufficiently long simulation time is taken.

\subsection{Temporal behavior}

\begin{figure}[tbp]
\centering
\includegraphics[width=8cm]{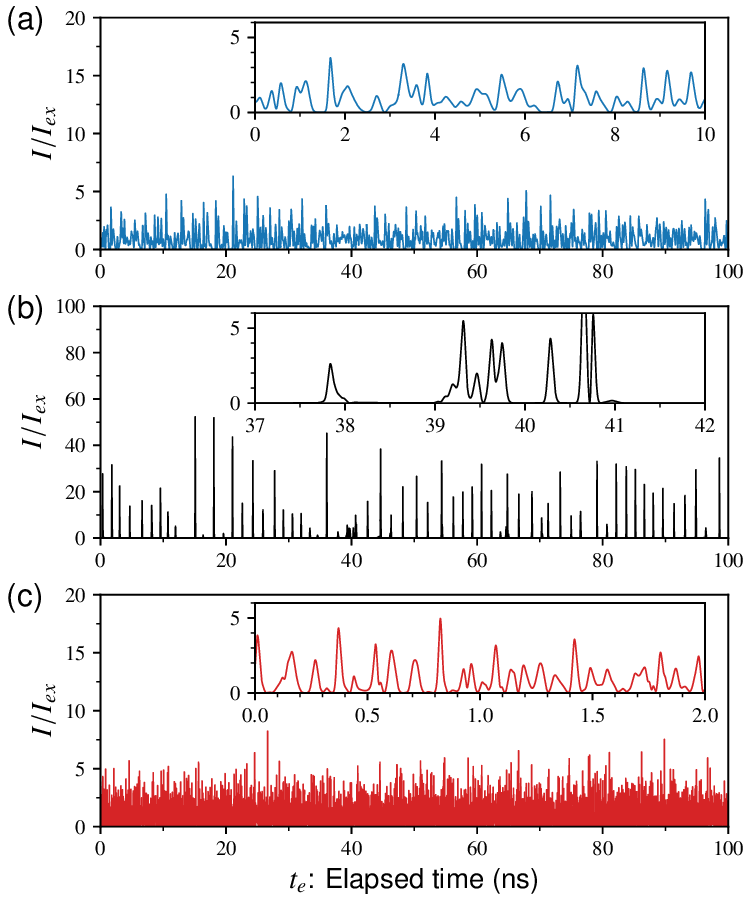}
 \caption{
Time series of normalized laser intensity 
obtained from the numerical simulations
under the condition of
$\tau = \SI{1}{\mu s}$ and $\kappa = \SI{60}{ns^{-1}}$, 
where the horizontal axis represents the elapsed time from 
$\SI{1999.5}{\mu s}$.
The normalized pump currents are (a) $j = 0.9579$, 
(b) $j = 1$, and (c) $j=3.5589$.
Irregular and intermittent pulsation (intensity bursts) with high intensity
is observed at the threshold pump current, $j=1$.
}
\label{fig:t-I-cri}
\end{figure}

Figure~\ref{fig:t-I-cri} shows 
the time series of normalized intensity $I(t)/I_{ex}$ obtained
from the simulation with a long delay time $\tau = \SI{1}{\mu s}$,
strong optical feedback $\kappa = \SI{60}{ns^{-1}}$, and
different pump currents ($j = 0.9579, 1$, and $3.5589$). 
The time series is shown as a function of the elapsed time
from $t = \SI{1999.5}{\mu s}$.
As shown in Fig.~\ref{fig:t-I-cri}(b),
a distinct behavior appears on the condition that the pump current
is just the threshold current, i.e., $j=1$;
irregular sharp pulses (intensity bursts) originate abruptly 
from a state of almost zero intensity.
On the contrary, as shown in Figs.~\ref{fig:t-I-cri}(a) and (c),
only chaotic fluctuation around a mean intensity, $I/I_{ex} \simeq 1$,
is observed when the currents that deviate from the threshold value
are applied ($j = 0.9579$ and $3.5589$).
It should be noted that the intensity rarely takes on a value near zero
when $j= 0.9579$ and $3.5589$.
This is in contrast to the case of $j=1$.
Here, we note that a sufficiently long simulation time 
is required to observe such intensity bursts shown in Fig.~\ref{fig:t-I-cri}(b)
because of the presence of a long transient regime in the early stage 
with quiescent intervals and quasi-periodic pulses.

Another significant feature is its scale-free burst-like pulsation behavior.
The peak heights of certain bursts attain several tens of times 
the steady intensity of the maximum gain mode, $I_{ex}$.
Recalling that $I_{ex}$ means the highest possible intensity produced by
steady-state oscillation,
we can consider that
the threshold condition provides a highly efficient laser oscillation.
On the other hand, significantly small bursts occur
as observed in the insert in Fig.~\ref{fig:t-I-cri}(b).
Although not apparent in the figure, 
there are many tiny pulses that are significantly smaller than $I_{ex}$.
As an example, one can find out a tiny pulse at $\SI{40.9}{ns}$ 
in the inset figure.

\subsection{Statistical distribution}

\begin{figure}[tbp]
\centering
\includegraphics[width=8.cm]{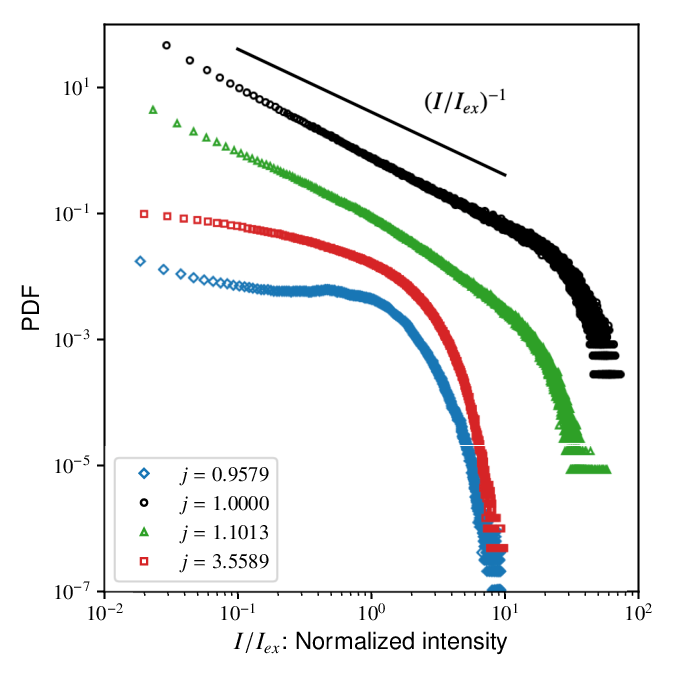}
 \caption{
Probability density function (PDF) of the normalized intensity $I/I_{ex}$
obtained from the numerical simulations, where
the PDFs are shifted arbitrarily in the vertical direction for visibility.
The distribution of the probability density of $j=1$ (black circles) 
clearly exhibits an algebraic decay with exponent $\beta \simeq 1$,
where the solid straight line represents $(I/I_{ex})^{-1}$.
The cutoff of the distributions is maximized at $j = 1$.
}
\label{fig:dist-I}
\end{figure}

To demonstrate the existence of 
scale-invariant fluctuations characterized by power-law distributions
in the laser oscillations,
we here present statistical distributions of the laser intensity
and intervals of the bursts calculated from the time series 
of the numerical simulations.

The probability density functions (PDFs) of the normalized intensity $I/I_{ex}$
are shown in Fig.~\ref{fig:dist-I}, where 
the probability distributions are depicted in a double logarithmic display.
Under the threshold condition ($j=1$), the distribution
evidently forms a straight shape (open black circles), 
i.e., the distribution contains a power-law decay described 
by Eq.~(\ref{eq:power-law}), over at least three orders of magnitude,
where the characteristic exponent is almost unity ($\beta \simeq 1$).
The exponent is close to that observed in seismic statistics
known as Gutenberg--Richter's law \cite{GutenbergRichter1956Law}.
When $j > 1$, 
the power-law decay (scale-invariant feature) 
disappears rapidly and transforms to an exponential one,
which has a characteristic scale.
When $j < 1$, the power-law feature also disappears,
as can be seen from the distribution represented by blue diamonds 
in Fig.~\ref{fig:dist-I}.
It should be noted that
the cutoff intensity of the distributions, i.e., 
the upper bound of the event size, 
takes on a maximum value under the threshold condition $j = 1$.
We will investigate this trend in detail in Subsec.~\ref{sec:condition}.

\begin{figure}[tbp]
\centering
\includegraphics[width=8cm]{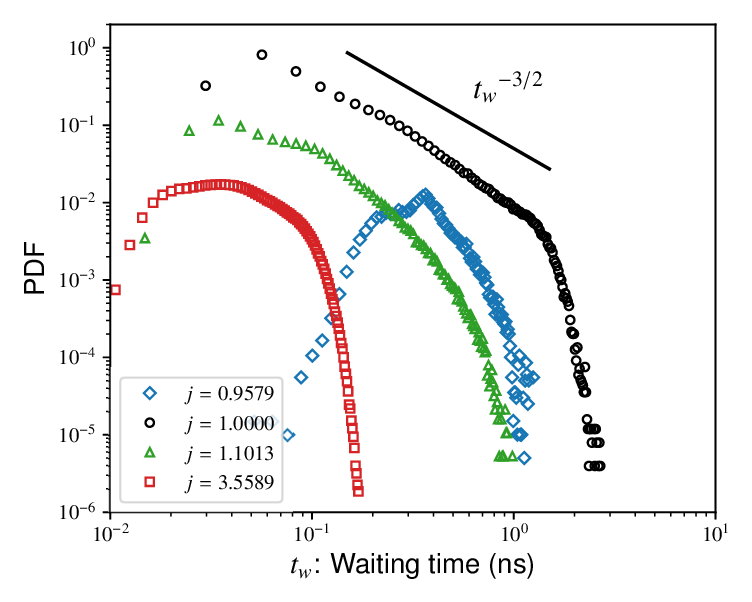}
 \caption{
PDF of waiting time,
the interval between successive peaks of intensity bursts,
obtained with an injection current $j$.
The power-law regime appears when $j = 1$ (black circles),
where the solid line represents a power-law curve with $\beta = 1.5$.
}
\label{fig:dist-t_w}
\end{figure}

It is known that critical behavior in nonequilibrium systems
yields power-law statistics
in terms of size and the time of energy-release events
\cite{Bak1987SOC,SOC1998Jensen,aschwanden2011SOC}.
Here, the distributions of waiting time, i.e., the time interval
between two sequential peaks of intensity bursts, 
are shown in Fig.~\ref{fig:dist-t_w},
where we consider each burst of intensity as an energy-release event.
The waiting time distribution obtained for a large pump current ($j=3.5589$) 
forms a convex shape around $t_w = \SI{0.04}{ns}$.
This means the existence of a characteristic timescale of the burst events.
With decreasing the pump current from the large value,
a straight ``shoulder,'' which is an indication of a power-law character,
develops in the distributions.
Under the threshold condition ($j=1$),
a power-law decay with $\beta \simeq 3/2$ clearly appears over approximately 
two orders of magnitude
as depicted by open black circles in Fig.~\ref{fig:dist-t_w}.
This power-law feature in the waiting time is compatible with
the aftershock statistics of earthquakes known as Utsu--Omori's law,
while the exponent of $3/2$ is slightly higher than that of the law 
\cite{Omori1894OmoriLaw,Utsu1969UtsuLaw}.
Similar to the case with the intensity statistics 
shown in Fig.~\ref{fig:dist-I},
the power-law behavior disappears again for lower pump currents ($j < 1$),
which are depicted by blue diamonds in Fig.~\ref{fig:dist-t_w}. 

The above simulation results show that
the statistical distributions of size and interval of energy-release events
form power-law decays just on the threshold condition $j=1$.
This indicates that semiconductor lasers with optical feedback
exhibit power-law statistics in the threshold condition of the solitary mode.

\subsection{Trajectory}

In this subsection, 
we observe the intensity dynamics exhibiting power-law fluctuations
as a trajectory in the phase space of time-delayed dynamical systems.
In the Lang--Kobayashi model, a space consisting of 
the phase difference $\Phi(t) = \phi(t - \tau) - \phi(t)$  
and carrier density $N(t)$ is called the ``phase space'' of the system. 
This space is only a projection of the rigorous phase-space.
However, it is commonly used to observe complex delay dynamics
such as LFFs and RPPs
\cite{Fischer1996LFF,Heil2001LK-RPP,Uchida2012ChaoticLaser},
because this phase space allows us to visually understand the relationship
between a trajectory and stationary points.
We here observe the phase space consisting of
the laser intensity $I(t)/I_{ex}$ rather than $N(t)$
since we have focused on the large intensity fluctuation in this study.

\begin{figure}[tbp]
\centering
\includegraphics[width=8cm]{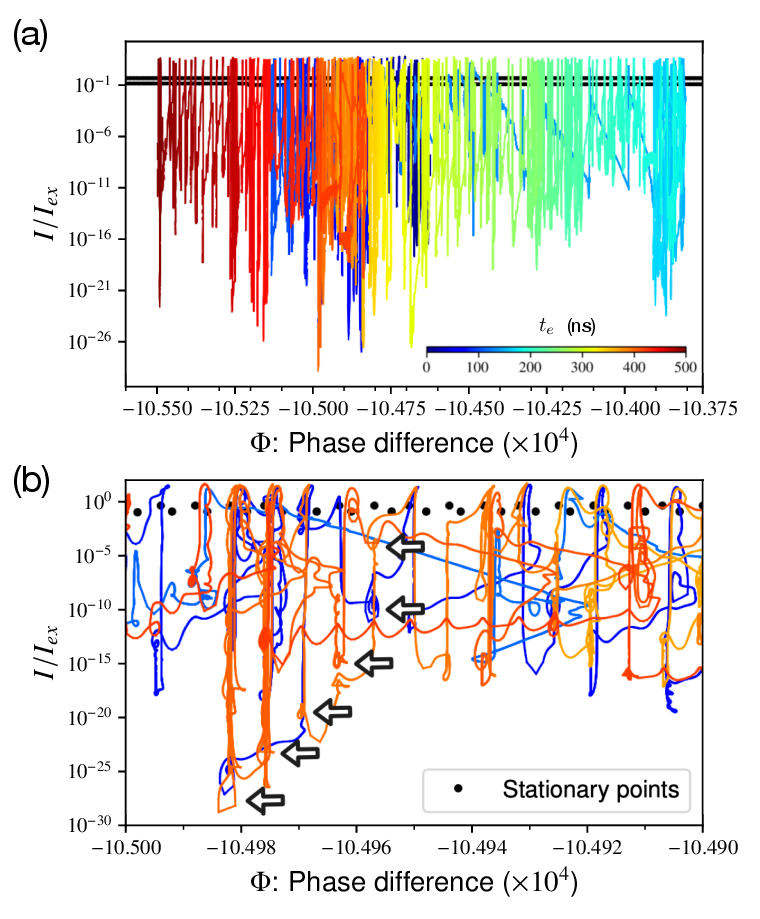}
 \caption{
A trajectory in the phase space constructed by phase difference 
$\Phi$ and normalized intensity $I/I_{ex}$ in the condition that
$\tau = \SI{1}{\mu s}$,
$\kappa = \SI{60}{ns^{-1}}$, and $j=1$, where the solid blue circles are
stationary solutions of the Lang--Kobayashi equation.
(a) The trajectory is depicted by color according to the elapsed time, $t_e$,
from $t = \SI{1999.5}{\mu s}$.
(b) Enlarged figure from $\Phi = -10.500$ to $-10.490$.
The white arrows indicate typical pseudo-stationary points (see text).
}
\label{fig:phase-space-cri}
\end{figure}

A part of a trajectory calculated from the numerical result 
accompanying the power-law behavior
is shown in Fig.~\ref{fig:phase-space-cri}.
Here, the simulation condition is identical to that used 
in Figs.~\ref{fig:t-I-cri}(b) and ~\ref{fig:dist-I}:
$\tau = \SI{1}{\mu s}$, $\kappa = \SI{60}{ns^{-1}}$, and $j=1$.
The curve in the figure represents the trajectory 
following $\SI{1999.5}{\mu s}$.
The blue solid circles are the stationary points of external cavity modes.
The figure shows certain unusual motions of the trajectory:
the trajectory moves largely 
omitting many stationary points when $\Phi(t)$ decreases
and increases abruptly in the direction of increasing $\Phi(t)$ 
(occasionally with a jerking motion).

The remarkably sticky motion of the trajectory is noteworthy;
The trajectory sometimes orbits several narrow regions
although there are no stationary points in the regions.
Some typical regions are indicated by white arrows in 
Fig.~\ref{fig:phase-space-cri}(b).
The narrow areas where the trajectory winds and/or slows down
can be regarded as points; hence, we refer to these points as
``pseudo-''stationary points.
The dynamics is more significantly affected by the pseudo-stationary points
than the ``ordinary-''stationary points depicted by the blue circles 
in the figure.
In LFFs and RPPs, trajectories follow a sequence from 
one ordinary-stationary point to the next, 
heading toward the maximum gain mode and returning to the minimum gain state 
after reaching the mode \cite{Fischer1996LFF,Heil2001LK-RPP,Uchida2012ChaoticLaser}.
In contrast, the trajectory of the present power-law dynamics 
is trapped by various pseudo-stationary points as well as 
the ordinary-stationary points and irregular jumps 
along the phase difference (in the horizontal direction).
This is an obvious difference between the present dynamics
and traditional complex dynamics, such as LFFs or RPPs.

Further, the pseudo-stationary points seem to be
(logarithmically) homogeneously distributed in a region that extends
from extremely small to large intensities in the phase space.
Thus, the power-law fluctuation is expected to extend to
a significantly smaller region of intensity
than that depicted in Fig.~\ref{fig:dist-I}.
The appearance of such multi-scale pseudo stationary points 
may be related to the power-law behavior in the intensity dynamics.
A more detailed analysis of the phase space dynamics 
would be reported elsewhere.

\subsection{Condition for power-law behavior}
\label{sec:condition}

\begin{figure}[tbp]
\centering
\includegraphics[width=7.cm]{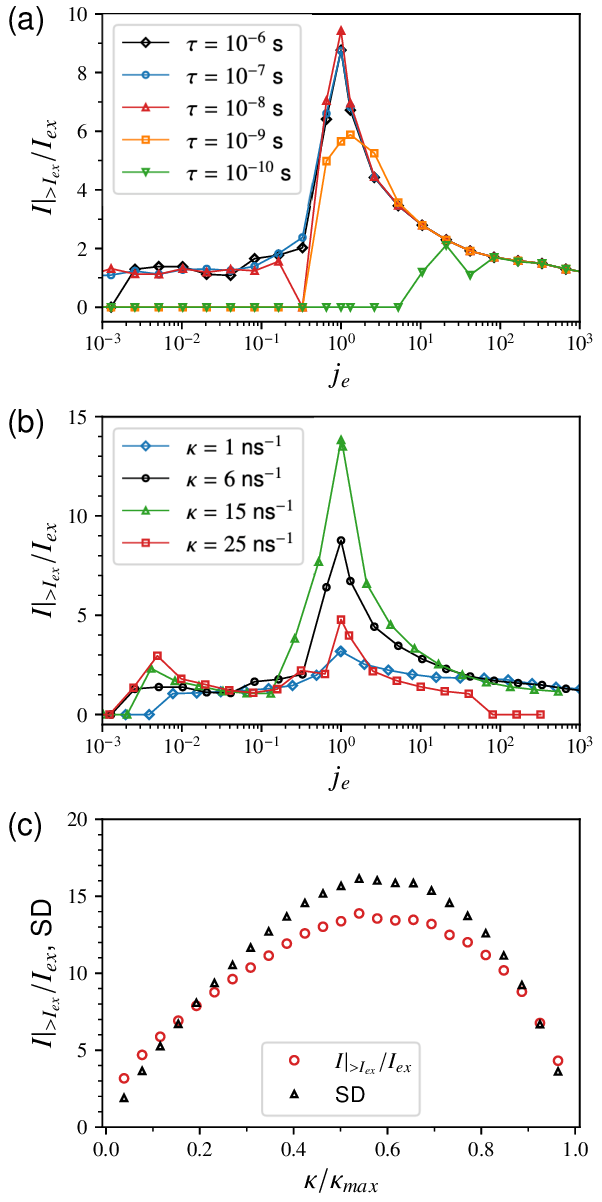}
 \caption{
Mean exceedance of the intensity, $\ave{I}|_{> I_{ex}}$, as a function of
the normalized pump current $j_e$ or 
feedback strength $\kappa/\kappa_{max}$, 
where $\ave{I}|_{> I_{ex}}$ is normalized by $I_{ex}$.
The mean values are obtained under the condition of
(a) various delay times $\tau$
and a common feedback strength $\kappa = \SI{60}{ns^{-1}}$,
(b) different $\kappa$ and a common delay time $\tau = \SI{1}{\mu s}$,
or (c) $\tau = \SI{1}{\mu s}$ and $j_e = 1$,
where the black triangles represent the standard deviation of 
$\ave{I}|_{> I_{ex}}/I_{ex}$.
}
\label{fig:j-I_ave}
\end{figure}

In this subsection, evaluating the extent of the power-law behaviors
systematically 
in various conditions of delay time $\tau$ and feedback strength $\kappa$,
we reveal the condition of the emergence of the power-law behaviors 
illustrated in the previous subsection.
To achieve this, 
we calculate the mean value of the {\it exceedance} of intensity,
$\ave{I}|_{>a}$, i.e., the size of the events beyond a reference value
$a$ \cite{Sornette2006SOC}, from the simulation results.
As shown in Fig.~\ref{fig:dist-I}, 
the distribution exhibits an evident power-law decay as the cutoff 
of the distribution increases toward the rightmost part of the figure.
Thus, the mean exceedance normalized by the maximum gain mode intensity,
$\ave{I}|_{> I_{ex}}/I_{ex}$,
can be a good indicator of the manifestation of power-law behaviors
\cite{Sornette2006SOC},
where we use $I_{ex}$ as the reference value for the exceedance calculation.
To represent the dependence of the pump current,
we employ another normalized pump current defined by
\begin{align}
 j_e \equiv \frac{J - J_{ex}}{J_{th} - J_{ex}} 
= \left( j - 1 \right) \frac{N_{th} G_N}{2 \kappa} + 1,
\end{align}
where $J_{ex}$ is the threshold pump current of the maximum gain mode
defined by $J_{ex} = J_{th} \left[ 1 - {2 \kappa}/(N_{th} G_N) \right]$
(see Appendix~\ref{sec:ext_mode}).
Thus, $j_e = 0$ and $j_e = 1$ 
correspond to the threshold currents of 
the maximum gain mode and solitary mode, respectively.

The dependence of the normalized mean exceedance $\ave{I}|_{> I_{ex}}/I_{ex}$ 
on the delay time $\tau$
is shown as a function of the pump current in Fig.~\ref{fig:j-I_ave}(a).
Here, we employed the time series after $t = \SI{1900}{\mu s}$ to calculate
the mean exceedance.
The figure shows a remarkable feature that 
$\ave{I}|_{> I_{ex}}/I_{ex}$ forms a sharp peak
around the threshold current of the solitary mode, $j_e = 1$, 
when the delay time is sufficiently long ($\tau > \SI{1}{ns}$).
This peak clearly represents the manifestation of the power-law behavior
at the threshold of the solitary mode.
The exceedance $\ave{I}|_{> I_{ex}}/I_{ex}$ approaches unity
as $j_e$ increases from the threshold current.
This indicates that the excess pump current changes
the power-law behavior accompanying extreme intensity bursts 
to steady oscillation with relatively small fluctuations.
On the other hand, $\ave{I}|_{> I_{ex}}/I_{ex}$ monotonically decreases 
as $j_e \to 0$ despite the divergence of $1/I_{ex}$ at $j_e = 0$.
This intensity disappearance implies that the maximum gain mode
no longer works at the threshold condition of the mode.

As shown in Fig.~\ref{fig:j-I_ave}(b),
similar peaks indicating the emergence of the power-law behavior
and the decay trend can be observed around $j_e = 1$,
but it is noteworthy that the peak disappears under 
an excessively large (or small) feedback strength condition.
This result implies the presence of 
a suitable feedback strength for the power-law behavior.
To identify the suitable condition, we depict the relationship between
 $\ave{I}|_{> I_{ex}}/I_{ex}$ with $j_e = 1$
and the normalized feedback strength $\kappa/\kappa_{max}$
in Fig.~\ref{fig:j-I_ave}(c),
where $\kappa_{max} = 1/(2 \tau_p) \simeq \SI{260}{ns^{-1}}$ which is 
the maximum value of the feedback strength as long as $I_{ex}$ is finite
[see Eqs.~(\ref{eq:I_max}) and (\ref{eq:kappa_max})].
In the figure,
the standard deviation depicted by open black triangles 
as well as $\ave{I}|_{> I_{ex}}/I_{ex}$
has a maximum value at $\kappa/\kappa_{max} \simeq 0.6$.

The results presented in this subsection reveal an emergence condition of 
the power-law feature: the laser oscillation with the power-law statistics
emerges under the condition that
the pump current is just on the threshold value of the solitary mode
and the delay time is sufficiently long ($\tau > \SI{1}{ns}$).
In addition, the power-law behavior is most remarkable
when the feedback strength is approximately 60 \% of the upper limit.

\section{Experimental Observation}
\label{sec:experiment}

In the previous section, we have demonstrated the power-law behavior
of laser intensity in the vicinity of the solitary mode 
with long time delay and strong feedback
by performing numerical simulations.
Here, we report the experimental results that support
the power-law behavior of the burst 
intensity dynamics presented in the previous section.

\subsection{Experimental setup}

\begin{figure}[tbp]
\centering
\includegraphics[width=8cm]{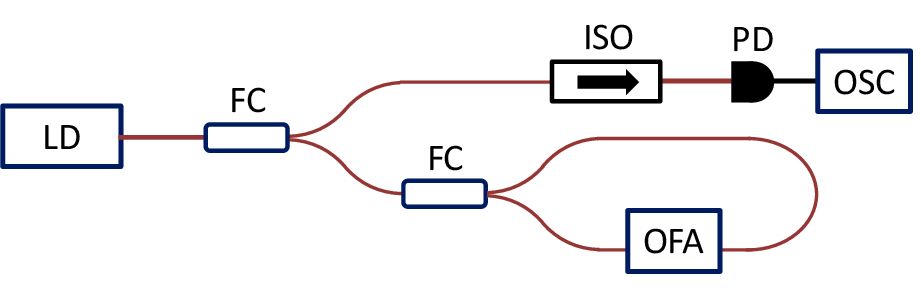}
 \caption{
Experimental setup for a semiconductor laser with delayed feedback.
LD, distributed feedback (DFB) laser diode; FC, 50/50 fiber coupler; 
ISO, optical isolator; PD, photodetector; OSC, oscilloscope; 
OFA, optical fiber amplifier.
The fiber loop is composed of polarization-maintaining fibers.  
}
\label{fig:exp-setup}
\end{figure}

Figure~\ref{fig:exp-setup} shows the experimental setup to 
measure the intensity time series of a semiconductor laser 
with delayed feedback. 
We used a distributed feedback (DFB) laser (operating at $\SI{1550}{nm}$) 
with a fiber feedback loop. 
The laser light is split into two portions by a 50/50 fiber coupler. 
One is sent to the photodetector (Bandwidth $\SI{12}{GHz}$) 
via an optical isolator, and the intensity signal is measured with a digital oscilloscope (Bandwidth $\SI{16}{GHz}$, 100 GSamples/s). 
The other is sent to an optical fiber amplifier 
and fed back to the DFB laser again. 
In this experiment, the laser current was fixed at $\SI{12.04}{mA}$ 
($\sim 1.2$ times of the threshold current) 
to measure the laser intensity signal clearly. 
The delay time was set as approximately $\SI{116}{ns}$,
which is sufficiently long to induce the power-law like behavior.  
We used the optical fiber amplifier to achieve a strong optical feedback,
and controlled the feedback ratio (represented by $R$) 
between the feedback power 
and output power from the laser to be within the range 
from $\SI{0}{\%}$ to $\SI{22}{\%}$. 

It should be noted that the power-law behavior is difficult to measure in the present experiment because of the limitation of the dynamic range of the measurement, i.e., the limitation of the measurement range of the 
8 bit analog-digital converter in the oscilloscope. 
Thus, the present experiment focus on the appearance of the intermittent burst of the intensity pulses, which is a signature of the power-law behavior reported in the previous section.

\begin{figure}[tbp]
\centering
\includegraphics[width=8cm]{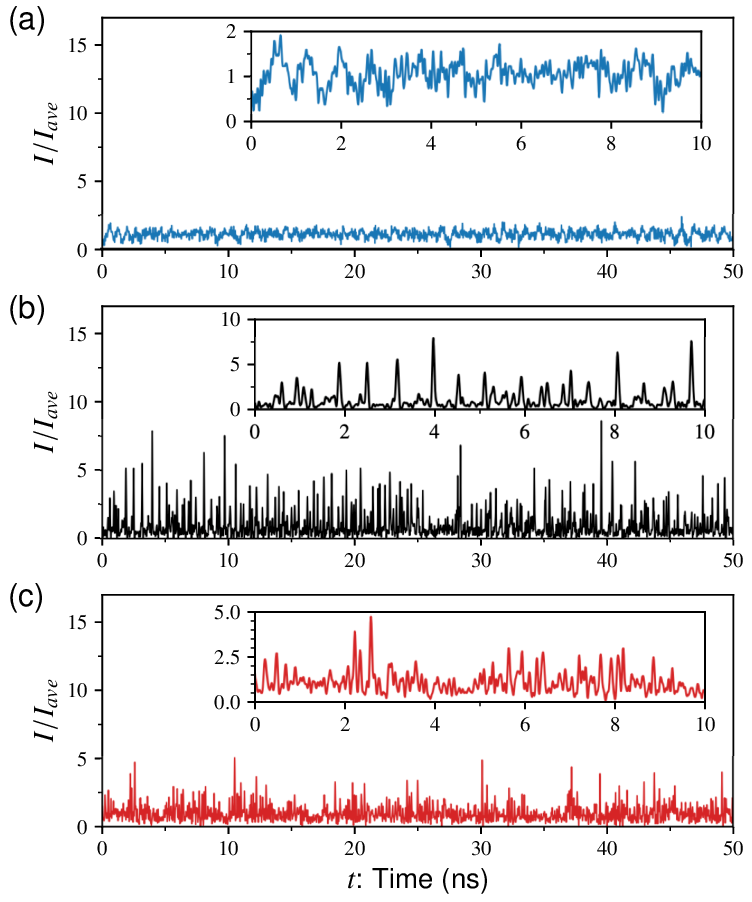}
 \caption{
Time series of light intensity at (a) $R =6.6 \%$, 
(b) $13 \%$ and (c) $22.5 \%$.
Enlarged time series are presented in each figure.
}
\label{fig:t-I-exp}
\end{figure}

\begin{figure}[tbp]
\centering
\includegraphics[width=7.5cm]{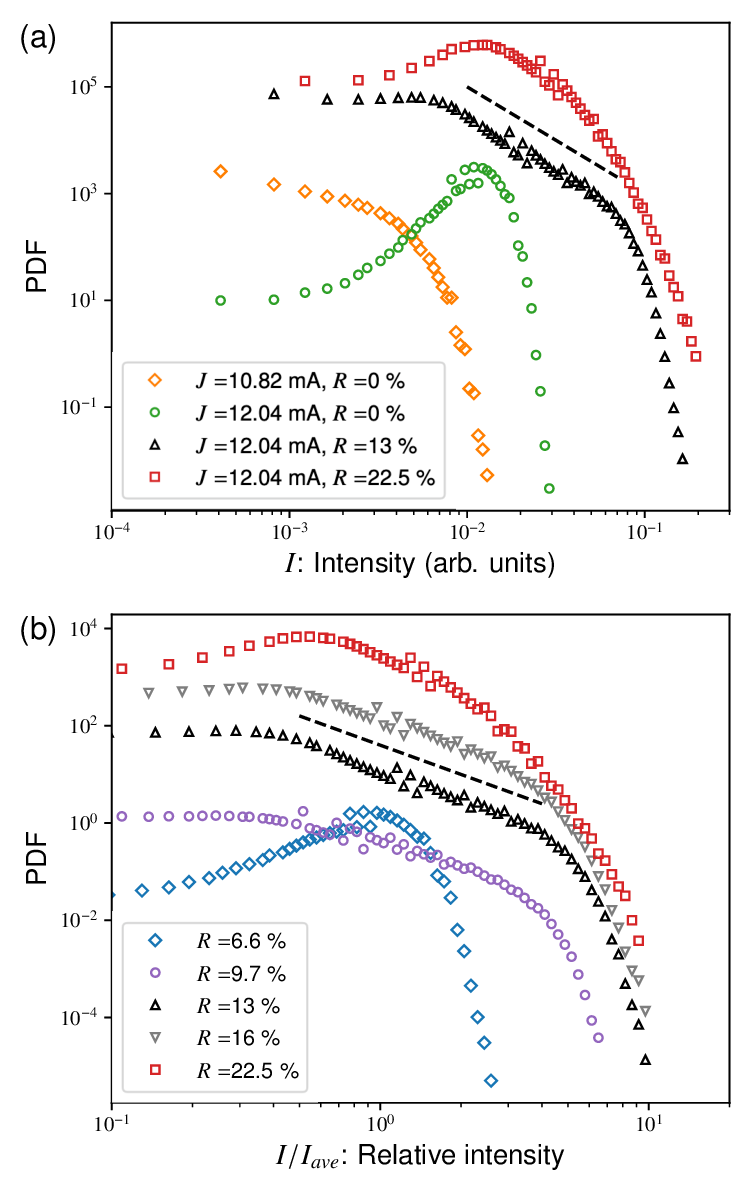}
 \caption{
Experimentally obtained PDF of intensity.
Here, all the plots are shifted arbitrarily in the vertical direction
for visibility.
(a) Intensity distributions obtained under the condition. 
The distribution of $J = \SI{10.82}{mA}$ and $R = \SI{0}{\%}$ 
is shown by orange diamonds,
where the laser oscillates did not occur.
The others are the distributions with $J = \SI{12.04}{mA}$ and 
different feedback strengths $R$.
(b) Intensity distribution normalized by the mean intensity $I_{ave}$.
All the pump currents are set as $J = \SI{12.04}{mA}$.
The dashed lines represent the power-law curve with the exponent $\beta = 2$.
}
\label{fig:dist-I-exp}
\end{figure}

\subsection{Temporal behavior}

Fig.~\ref{fig:t-I-exp} shows typical examples of 
intensity time series of laser oscillation obtained from our experiments at 
$J= \SI{12.04}{mA}$.
The time series at a low feedback-power ratio of $R = \SI{6.6}{\%}$
does not indicate a significant burst [Fig.~\ref{fig:t-I-exp}(a)].
On the other hand, irregular intensity bursts emerge when $R$ increases 
[Fig.~\ref{fig:t-I-exp}(b)]
and become inactive for a large feedback ratio of 
$R = \SI{22.5}{\%}$, 
where the intensity fluctuates around a mean value. 
This dependence on feedback strength is consistent with 
our simulation results depicted in Figs~\ref{fig:j-I_ave}(b) and (c).

\subsection{Statistical distribution}

Probability density functions of the intensity obtained from
the experimental time series at certain typical conditions
are shown in Fig.~\ref{fig:dist-I-exp}(a),
where the time series until $\SI{40}{\mu s}$ is used for the calculation.
In the figure, the distributions are shifted arbitrarily 
in the vertical direction for visibility.

A sufficient pump current for laser oscillation without delayed feedback 
($J= \SI{12.04}{mA}$, $R= \SI{0}{\%}$)
yields only a Gaussian-like distribution,
which is depicted as green circles in the figure.
This reflects the fluctuation behavior near the mean intensity. 
When the optical feedback is applied,
the fluctuation is increased dramatically, and 
the regime of power-law-like decay appears in the statistical distribution.
As shown by the black triangles 
in Fig.~\ref{fig:dist-I-exp}(a),  
the power-law regime is indicated by the dashed line in the figure,
and the characteristic exponent is approximately two ($\beta \simeq 2$).
This experimental condition ($J= \SI{12.04}{mA}$, $R= \SI{13}{\%}$) 
is identical to that of Fig.~\ref{fig:t-I-exp}(b).

As our numerical simulations predict that
this power-law behavior disappears when the feedback strength 
is significantly larger,
the same tendency can be observed in the present experiments.
Fig.~\ref{fig:dist-I-exp}(b) shows statistical distributions of 
relative intensity under the fixed pump current ($J= \SI{12.04}{mA}$)
and different feedback strengths ($R = 6.6, 9.7, 13$, and $\SI{22.5}{\%}$),
where the intensity is normalized by the mean intensity $I_{\text{ave}}$.
The figure reveals that
the power-law regime becomes prominent in the distribution
when the feedback strength increases,
but becomes less prominent when the strength exceeds $\SI{13}{\%}$.
The cutoff of the distributions,
i.e., the maximum relative intensity, also has 
a character similar to that of the numerical results 
as shown in Fig.~\ref{fig:dist-I}:
the cutoff becomes maximum at $R = \SI{13}{\%}$ 
even the difference among the cutoffs is also not clear.

\section{Discussion}
\label{sec:discussion}

This study shows that intermittent intensity bursts
characterized by power-law statistics 
emerge and are prominent when the delayed feedback is
sufficiently long and strong
and the pump current is just on the lasing threshold.
In order to explain the power-law behavior, 
as a starting point for the discussion,
it would be natural to focus on 
the nonequilibrium-critical phenomena
in slow driven-dissipative systems known as self-organized criticality (SOC),
a mechanism that produces
similar power-law distributions \cite{Bak1987SOC,Bak1988SOC}.
SOC is a phenomenon that a dissipative system spontaneously organizes itself 
to a critical state that is characterized by scale-invariant statistics,
i.e., power-law distribution.
It is considered that the emergence of the criticality requires
several conditions for a system that 
stores the energy injected by external driving \cite{SOC1998Jensen}:
(i) the system dissipates the stored energy through 
{\it multiple} energy-release processes. 
(ii) the processes interact with each other,
and (iii) the processes have threshold values for activation,
and (iv) the rate of energy injection into the system 
is sufficiently slower than that of the energy-release processes.

The external energy injection of the laser system considered in this study
is the pump current $j$.
In laser systems without feedback,
the slowest driving rate corresponds to the threshold current ($j = 1$)
rather than zero pump current ($j=0$).
This is because the carrier density of the laser decays with lifetime $\tau_s$
even in the absence of energy release by laser oscillation
[see Eq.~(\ref{eq:LK-complex-N})].
Thus, the threshold current can be regarded as 
the ideal situation that satisfies the fourth condition.
The number of virtual nodes of a time-delay dynamical system is proportional to
the delay time $\tau$ \cite{Uchida2012ChaoticLaser}.
Furthermore, the strength of the interaction between 
the feedback and laser intensity is determined by 
the coupling strength of the feedback, $\kappa$.
Hence, a sufficiently long $\tau$ is required for 
the multiplicity (high dimensionality) of the energy release process,
and a strong $\kappa$ firmly establishes their interactions.
Therefore, it is reasonable that the power-law fluctuation 
is prominent in the situation.
However, certain results cannot be explained directly 
by the occurrence conditions of SOC.
As shown in Figs.~\ref{fig:j-I_ave}(a) and (b),
the power-law behavior is prominent when $j_e = 1$ rather than when $j_e = 0$.
It is a wonder that the onset of the power-law behavior
is determined by the threshold of the solitary mode $j_e = 1$
rather than that of the external cavity modes $j_e = 0$,
although the power-law behaviors depend strongly on the external 
feedback light.
In future works, more efforts should be undertaken to explain
the optimum value of the feedback strength, 
$\kappa/\kappa_{max} \simeq 0.6$, 
to obtain remarkable power-law fluctuations 
as shown in Fig.~\ref{fig:j-I_ave}(c).

At a glance, the aforementioned conditions, particularly 
the requirement of the injection current tuning, appear to 
deviate from the concept of ``self-organized'' criticality.
This is because parameters governing the conditions exist clearly 
in laser systems, whereas they are presumed implicitly 
in conventional SOC systems.
For instance, forest fires or earthquakes, 
known as typical examples of SOC systems, seem to spontaneously organize 
a critical state without any fine-tuning of parameters.
The models of those phenomena have no explicit parameters 
for controlling energy injection rates 
as they introduce an infinite timescale separation a priori. 
This separation is attributed to the fact that the injection rates, 
such as tree growth and stress loading on a fault 
by the motion of the tectonic plates,
are extremely slow compared to the energy release
by fire spread in a forest and fault slip, respectively.
The influence of the timescale separation on the criticality 
was discussed by Vespignani and his coworkers, 
introducing a nonvanishing driving rate into the sandpile system, 
which is a traditional toy model of SOC \cite{Vespignani1998SOCMFT}.
By contrast, laser systems innately contain a parameter 
for naturally controlling the driving rate as the injection current.
Hence, the laser with delayed feedback is an extraordinary example of SOC 
and can be a useful instance of embodying the effect of timescale separation 
on criticality.

Resonance often plays a key role in irregular laser oscillations,
but the resonance is not expected to contribute to the power-law behavior 
in the present condition.
The resonance considered here results from the interaction 
between the solitary mode,
which is the intrinsic oscillation mode of the semiconductor lasers,
and the optical feedback with the time delay $\tau$.
It is well known that the frequency of this solitary mode is described by 
$f_r = {1}/{(2 \pi)} \sqrt{{(j-1)}/{(\tau_s \tau_p)} (1 + G_N N_0 \tau_p)}$
\cite{Uchida2012ChaoticLaser},
but the frequency is zero under the threshold condition ($j=1$).
Thus, the resonance with the solitary mode cannot occur
for the threshold condition considered in the present study
\cite{Liu1995delayedLaser}.
The absence of the frequency is also consistent with 
the power-law decay of the waiting time distribution at the threshold current
depicted in Fig.~\ref{fig:dist-t_w}, where the decay indicates that
the characteristic timescale of the bursts is absent.

Next, we mention the discordance between the experimental and numerical results.
The width of the power-law regime and the characteristic exponent
obtained in our experiments
vary slightly from the simulation results.
The power-law regime obtained in the experiments is limited to one order of
magnitude [from $I = 0.01$ to $0.1$ in Fig.~\ref{fig:dist-I-exp}(a)]
unlike the numerical results (Fig.~\ref{fig:dist-I}).
The lower bound would result from the noise of the measurement,
whose size can be measured as the distribution 
in the non-oscillatory condition depicted by orange diamonds
in Fig.~\ref{fig:dist-I-exp}(a).
This noise region is almost consistent with the lower bound 
of the power-law regime of the experiments.
The dynamic range of the measurement equipment used in the experiments
is also a reason for the lower bound.
The large characteristic exponent $\beta \simeq 2$ in the experiment
compared with that in the simulations may have originated from 
the gain saturation of the laser and incomplete tuning of the pump current
to the threshold value.
The saturation effect reduces the cutoff 
of the intensity distributions (see Appendix~\ref{sec:saturation}).
The combination of the saturation effect and the excess pump current
due to the incomplete tuning 
causes a partial increase in the gradient of the intensity 
distribution to $1.8$ (Fig.~\ref{fig:dist-I-satu}).

Here, we indicate that the burst behavior is different from
previously reported dynamical behaviors 
such as chaotic oscillations, RPPs, and LFFs.
The pulses that construct the bursts may appear to be similar to RPPs,
but, unlike RPPs,
periodicity of the pulses 
is absent \cite{Heil2001LK-RPP,Uchida2012ChaoticLaser}.
Even though the regular dropouts peculiar to LFFs are not observed 
in the burst behavior \cite{Fischer1996LFF,Heil1998LFFmap,Ohtsubo2012semiconductor,Uchida2012ChaoticLaser},
LFFs may be associated with the power-law behaviors.
This is because the condition for the emergence of the power-law bursts 
is close to that for LFFs except for the delay time. 
According to our investigation, depending on the feedback strength,
regular drops can be observed when $\tau_D \lesssim \SI{0.05}{ns}$,
and a longer delay time causes the time series to be irregular.
Future studies would elucidate 
the connection between the present power-law bursts and LFFs.

\begin{figure}[tbp]
\centering
\includegraphics[width=7.5cm]{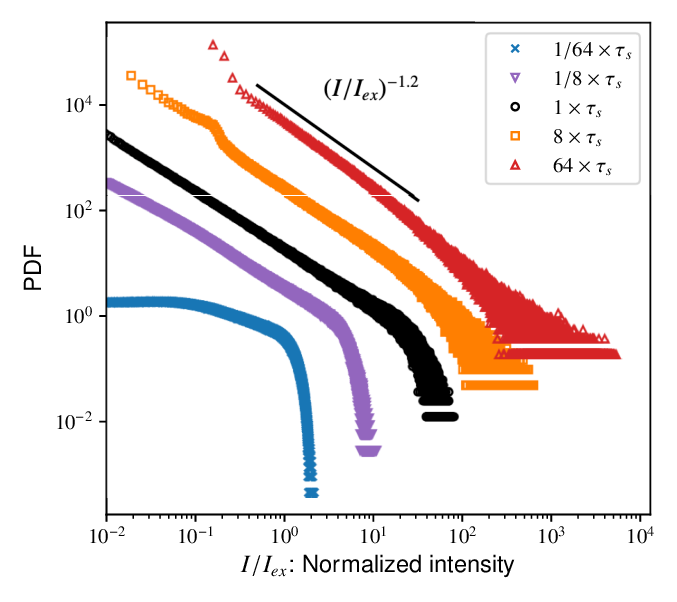}
 \caption{
The intensity distribution with the different carrier lifetime $\tilde{\tau}_s$.
The distributions become to loose their cutoff and exhibit a simple
power-law decay as $\tilde{\tau}_s$ is longer.
}
\label{fig:dist-I-tau_s}
\end{figure}

\begin{figure}[tbp]
\centering
\includegraphics[width=7.5cm]{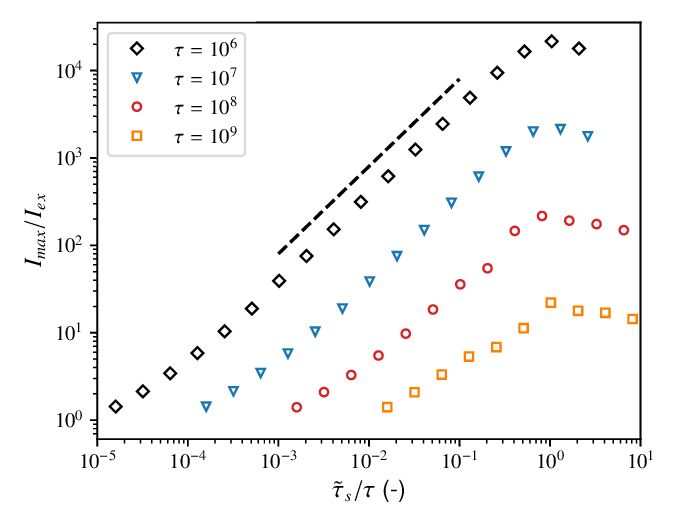}
 \caption{
Relative maximum intensity 
as a function of carrier lifetime $\tilde{\tau}_s$, where the lifetime
is normalized by delay time $\tau$.
The maximum intensity increases with increasing $\tilde{\tau}_s$,
but the trend disappears as $\tilde{\tau}_s$ exceeds $\tau$,
where the dashed line represents the linear relation, 
$I_{\mathrm{max}}/I_{\mathrm{ex}}\sim \tilde{\tau}_s/\tau$.
}
\label{fig:tau_s-depend}
\end{figure}


Finally, we discuss another factor that enhances the power-law behavior
and its relationship with SOC.
Even under the threshold condition with a sufficiently 
strong and long feedback, the intensity distribution still exhibits
a sharp cutoff that restricts larger events, as shown in Fig.~\ref{fig:dist-I}.
According to the mean-field theory of SOC,
dissipation in bulk and boundary regions of a system 
is essential for the criticality of SOC 
\cite{Vespignani1997SOCMFT,Vespignani1998SOCMFT}.
As semiconductor laser systems have no boundaries, 
here, we focus on the bulk dissipation, i.e., intrinsic dissipation.
The intrinsic dissipation of semiconductor laser system
is the spontaneous annihilation of carriers, whose timescale is
determined by the carrier lifetime $\tau_s$.
Though, the lifetime is a specific constant of the laser,
it enables a change in the value in the numerical simulations.
Figure~\ref{fig:dist-I-tau_s} illustrates the intensity distribution
obtained through simulations with different carrier lifetimes:
$\tilde{\tau}_s = \tau_s/64$, $\tau_s/8$, $\tau_s$, $8 \tau_s$, and $64 \tau_s$.
As shown in the figure, 
a long carrier lifetime (i.e., small intrinsic dissipation)
makes the cutoff of the distributions large
and the distributions approach a pure power-law distribution.
This is consistent with the conclusion of the mean-field theory of SOC 
\cite{Vespignani1997SOCMFT,Vespignani1998SOCMFT}.

When carrier lifetime is too long, however, the criticality characterized
using the cutoff of the power-law distribution breaks.
We show the maximum intensity values as a function of 
$\tilde{\tau}_s$ normalized by the delay time $\tau$ 
in Fig.~\ref{fig:tau_s-depend}.
The maximum intensity increases almost linearly. However, this trend
breaks down in a regime where $\tilde{\tau}$ is close to $\tau$.
The delay time must be sufficiently longer than the carrier lifetime
to achieve the power-law behavior.
Therefore, the criticality of this delay system is expected to be 
attained at the limits where
$\tilde{\tau}_s \to \infty$, $\tau \to \infty$, where $\tilde{\tau}_s < \tau$.

Nonequilibrium critical phenomena, namely SOC mentioned here,
are closely related to nonequilibrium phase transitions, 
particularly the absorbing-state phase transition
\cite{Munoz1999ExpoAbsStatTrans,Vespignani2000ASPTSOC,Henkel2008nonEquiPhaseTrans}.
In fact, absorbing-state phase transition has been observed in
a semiconductor laser with delayed feedback 
similar to that in the present study \cite{Faggian2018LaserAbsorbTrans}.
It is also noteworthy that the dynamical systems with time delays
can produce SOC-like behaviors,
in contrast to the fact that the conventional models that reproduce SOC and
absorbing-state phase transitions generally consist of cellular automata
\cite{Bak1987SOC,Bak1988SOC,SOC1998Jensen,aschwanden2011SOC}.
The manner in which the delay system produces the power-law behaviors 
can be observed as a trajectory in the phase space as depicted 
in Fig.~\ref{fig:phase-space-cri}.
Since most of the discussion on the relationship between the present
results and SOC is conjectural,
the detailed mechanism that produces the SOC features 
in delay systems should be elucidated in future studies.
It might enable the understanding of nonequilibrium critical phenomena
and phase transitions from the aspect of dynamical systems 
in the presence of delayed feedback.

\section{Conclusion}
\label{sec:conclusion}

In the present simulations and experiments,
we demonstrated that the semiconductor lasers with a sufficiently strong 
and long delayed feedback at the critical point (lasing threshold)
can cause the intermittent intensity fluctuation 
characterized by power-law distributions 
typical in nonequilibrium critical phenomena.
The results of the numerical simulations of the Lang--Kobayashi model
show that the laser intensity exhibits irregular bursts consisting of 
spike-like pulsation.
As a result, certain peak values of the bursts attain several tens of times 
the intensity of the maximum gain mode.
This is quantitatively different from other currently reported behaviors
such as low-frequency fluctuations, 
regular pulse packages, or chaotic oscillations.
Similar behaviors were experimentally measured in a semiconductor laser
with delayed feedback. 
The intensity bursts are most remarkable when
the coupling of the feedback is moderately strong 
($60 \%$ of the maximum feedback strength),
the delay time is sufficiently long ($\tau > \SI{1}{ns}$), 
and the pump current is just on the threshold value of the solitary mode.
The noteworthy feature is that the intensity and waiting time of the bursts 
follow power-law distributions with exponents of 
approximately $1$ and $2/3$, respectively.
Additional simulations indicate another non-trivial condition 
for the power-law behavior;
The behavior is more remarkable 
when the carrier lifetime of the laser increases
and the lifetime is shorter than the delay time.
The statistical feature and the condition of the intermittent laser oscillation
are analogous to self-organized criticality.

\begin{acknowledgments}
The author S.~S. thanks Dr. Kenichi Arai 
(NTT Communication Science Laboratories) for his support 
for the laser experiment.
This work was partly supported by JSPS KAKENHI 
(Grant No.~20K03783 and ~20H042655) and JST PRESTO (Grant No.~JPMJPR19M4). 
\end{acknowledgments}

\appendix

\section{Characteristic values of external cavity modes}
\label{sec:ext_mode}

To investigate the intensity dynamics of the lasers,
in this research, we employed certain quantities that represent external cavity
modes of semiconductor lasers described by the Lang--Kobayashi equations;
the oscillation intensity of maximum gain mode $I_{ex}$,
threshold current of the external cavity modes $J_{ex}$, 
and maximum strength of the delayed feedback $\kappa_{max}$.
Although these 
are described in commonly available textbooks \cite{Uchida2012ChaoticLaser},
we introduce these quantities in this appendix
for the reader's convenience.

The intensity generated from the maximum gain mode can be estimated
by considering the stationary state of the Lang--Kobayashi equation:
$\pdev{\hat{E}}{t}{} = \pdev{N}{t}{} = 0$.
The stationary carrier density $N_s$ when $\pdev{\hat{E}}{t}{} = 0$
is obtained from Eq.~(\ref{eq:LK-complex-E}):
\begin{align}
  N_s = N_{th} - \frac{2 \kappa}{G_N} \cos \omega_s \tau
\end{align}
where $N_{th} = N_0 + {1}/({\tau_p G_0})$, and
$\omega_s$ is a stationary angular frequency.
Substituting the carrier density into Eq.~(\ref{eq:LK-complex-N}),
we can obtain 
the stationary intensity of external cavity modes described
by the square of $\hat{E}$:
\begin{align}
I_s = 
\frac{\tau_p}{G_N \tau_s } 
\frac{(j-1) N_{th} G_N + 2 \kappa \cos \omega_s \tau}
{ 1 - 2 \tau_p \kappa  \cos \omega_s \tau },
\label{eq:I_s_multi} 
\end{align}
where $j$ is the normalized pump current defined by $j = J / J_{th}$,
and $J_{th} =  N_{th}/ \tau_s$.
Because $I_s$ is determined by the angular frequency of a mode $\omega_s$,
it is maximized when $\cos \omega_s \tau = 1$.
Hence, the intensity of the maximum gain mode that we used to normalize
the intensity of oscillation in this study can be represented by
\begin{align}
  I_{ex} = \frac{1}{G_N} \frac{\tau_p}{\tau_s} 
\frac{ (j-1) N_{th} G_N + 2 \kappa }{1 - 2 \tau_p \kappa}.
\label{eq:I_max-appendix}
\end{align}

The condition $I_{ex} = 0$ means that
at least one external cavity mode begins to oscillate.
Thus, the external cavity mode can activate at the normalized pump current
$j = 1 - {2 \kappa}/({ N_{th} G_N})$.
Then the threshold pump current is represented by 
\begin{align}
J_{ex} = J_{th} \left( 1 - \frac{2 \kappa}{N_{th} G_N} \right).
\label{eq:j_c}
\end{align}

From the denominator of Eq.~(\ref{eq:I_max-appendix}),
we can recognize that the stationary intensity is finite unless 
$2 \tau_p \kappa \ge 1$.
Therefore, the maximum feedback strength that we can set is determined by
\begin{align}
 \kappa_{max} = (2 \tau_p)^{-1}.
\label{eq:kappa_max}
\end{align}
These estimations are good approximations as long as the number of
external cavity modes is sufficiently large.

\section{Saturation effect of gain}
\label{sec:saturation}

\begin{figure}[tbp]
\centering
\includegraphics[width=8cm]{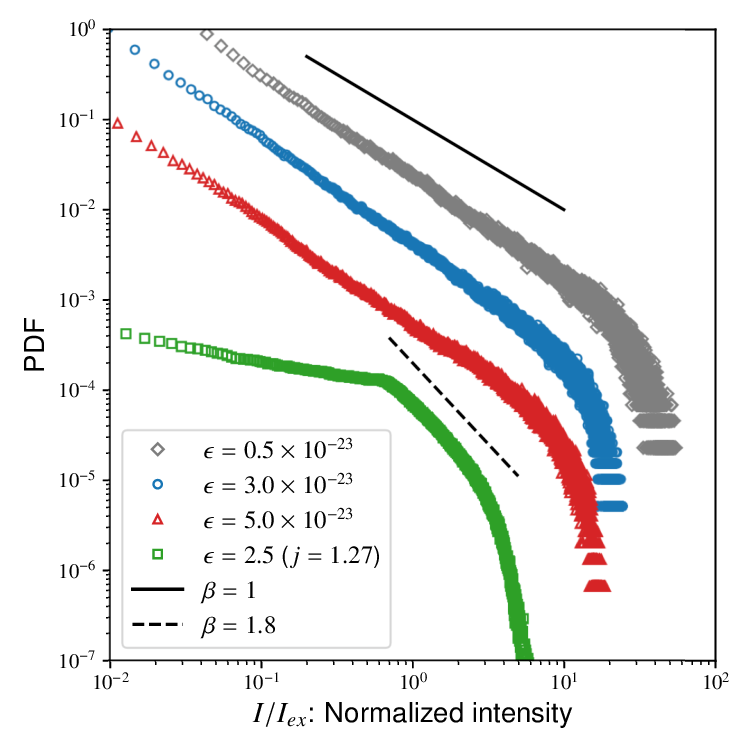}
 \caption{
PDF of oscillation intensity for different gain saturation parameters
$\epsilon$ with the threshold injection current $j=1$,
where the distribution depicted by green squares is for
$\epsilon = \SI{2.5e-23}{m^3}$ and $j = 1.2652$.
For visibility, power-law curves with exponents $\beta = 1$ and $1.8$
are shown by solid and dashed lines, respectively.
}
\label{fig:dist-I-satu}
\end{figure}

\begin{figure}[tbp]
\centering
\includegraphics[width=8cm]{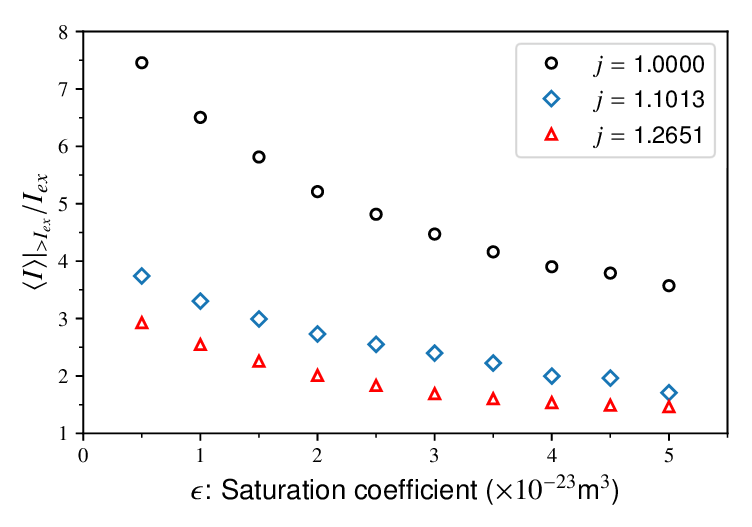}
 \caption{
The mean exceedance of the intensity for different injection currents $j$
as a function of the saturation parameter $\epsilon$,
where the exceedance is normalized by $I_{ex}$.
}
\label{fig:eps-I_ave}
\end{figure}

To extract the essence of power-law behaviors in semiconductor
lasers with delayed feedback, we omitted the effect of gain saturation 
of the laser in the simulations of this study.
However, the saturation effect 
is essential for comparing the present simulations with the experiments.
Here, we show the influence of the gain saturation upon the intensity
distributions by performing simulations of the Lang--Kobayashi model
with the nonvanishing gain saturation parameter $\epsilon$ 
shown in Eq.~(\ref{eq:LK-G_N}).
For the simulations, we applied some values of the saturation parameter
in the range from $\epsilon = \SI{0.1e-23}{m^3}$ to $\SI{5.0e-23}{m^3}$,
where the range includes the value of actual lasers used in 
ordinary experiments 
\cite{Harayama2011ChaosLaserRandomGen, Mikami2012EntropyPhysRand}.

The intensity distributions with three gain saturation parameters
are shown in Fig.~\ref{fig:dist-I-satu}, where 
the gray diamonds, blue circles, and red triangles correspond to
$\epsilon = \SI{0.5e-23}{}$, $\SI{3.0e-23}{}$, and $\SI{0.5e-23}{m^3}$,
respectively, and the pump current of the three distribution
is set as the threshold value of the solitary mode ($j=1$).
As is evident from the figure,
the results of $\epsilon = \SI{0.5e-23}{}$ and $\SI{3.0e-23}{m^3}$
exhibit power-law distributions with $\beta \simeq 1$
that are identical to those obtained by the simulations with $\epsilon = 0$.
The larger saturation parameter ($\epsilon = \SI{5.0e-23}{m^3}$)
affects its distribution slightly.

A more significant influence of the gain saturation is a decrease in 
the cutoff of the distributions with an increase in $\epsilon$.
This trend can be confirmed more clearly in Fig.~\ref{fig:eps-I_ave},
where the mean exceedance above $I_{ex}$ is plotted as a function of $\epsilon$.
The figure indicates that the maximum intensity caused by oscillation bursts
decreases monotonically with $\epsilon$ regardless of $j$.
Another notable result can be observed in the distribution 
as shown by green squares in Fig.~\ref{fig:dist-I-satu}.
Here, its saturation parameter, $\epsilon = \SI{2.5e-23}{m^3}$, 
is comparable with that used for 
the present experiments described in Sec.~\ref{sec:experiment}, 
and the pump current is slightly 
larger than the solitary mode threshold, $j = 1.2652$.
The distribution indicates that
a synergy between the gain saturation and the excess of the current
disrupts a wide range of power-law regimes, and
the gradient of the power-law regime of the distribution increases
more than $\beta = 1.8$.
This implies that power-law statistics different from that obtained 
in the simulations can be observed in the experimental measurements
because it is hard to tune the injection current just to the threshold
value of the solitary mode
and the saturation effect is consistently present in actual lasers.

The decrease in the cutoff of the intensity distribution 
due to the gain-saturation effect 
can be explained in terms of the SOC framework.
Gain saturation is considered to be a type of dissipation in laser systems.
As discussed through mean-field theory 
\cite{Vespignani1997SOCMFT,Vespignani1998SOCMFT}, 
intrinsic dissipation and/or the finite size effect weaken 
the criticality (see also the dependence of the carrier lifetime 
in Sec.~\ref{sec:discussion}). 
Thus, the shortening the width of the scaling region by introducing 
the gain saturation can be understood as the influence of such effects.

\bibliographystyle{apsrev}
\bibliography{NECB-Laser}

\end{document}